\documentclass[aps,prl,amsmath,amssymb,floatfix,lengthcheck,reprint]{revtex4-1}%{article}%
\usepackage{graphicx}
\begin{document}
\title{Time-Domain Detection of Weakly Coupled TLS Flucuators in Phase Qubits}
\author{M.S. Allman, F. Altomare, J.D. Whittaker, K. Cicak, D. Li, A. Sirois, J. Strong, J.D. Teufel, R.W. Simmonds}
\email{simmonds@boulder.nist.gov}
\address{National Institute of Standards and Technology, 325 Broadway, Boulder, Colorado 80305-3328, USA}
\date{\today}

\begin{abstract}
We report on a method for detecting weakly coupled spurious
two-level system fluctuators (TLSs) in superconducting qubits.
This method is more sensitive that standard spectroscopic
techniques for locating TLSs with a reduced data acquisition time.
\end{abstract}

\maketitle

Superconducting qubits are showing promise as viable candidates
for implementing quantum information processing\cite{Clarke,
MartinisB}.  However, spurious two-level system fluctuators (TLSs)
are still believed to be a major source of decoherence in phase
qubits\cite{Simmonds,MartinisC}.  Spectroscopic measurements are
the traditional means of locating TLSs associated with defects in
the tunnel barrier of the qubit's Josephson
junction\cite{Simmonds}. Saturation effects from long excitation
pulses and relatively broad qubit linewidths ($\sim 2-10\,MHz$)
can prevent weakly coupled or weakly coherent TLSs from being
visible with standard spectroscopic measurements.  We report on a
time-domain method for resolving weakly coupled TLS junction
fluctuators that is more sensitive than standard spectroscopic
techniques, resolving fluctuators with coupling strengths below
$10\,MHz$, and with considerably shorter acquisition times.

\begin{figure}[!htbp]
\centering
\includegraphics[width=\columnwidth]{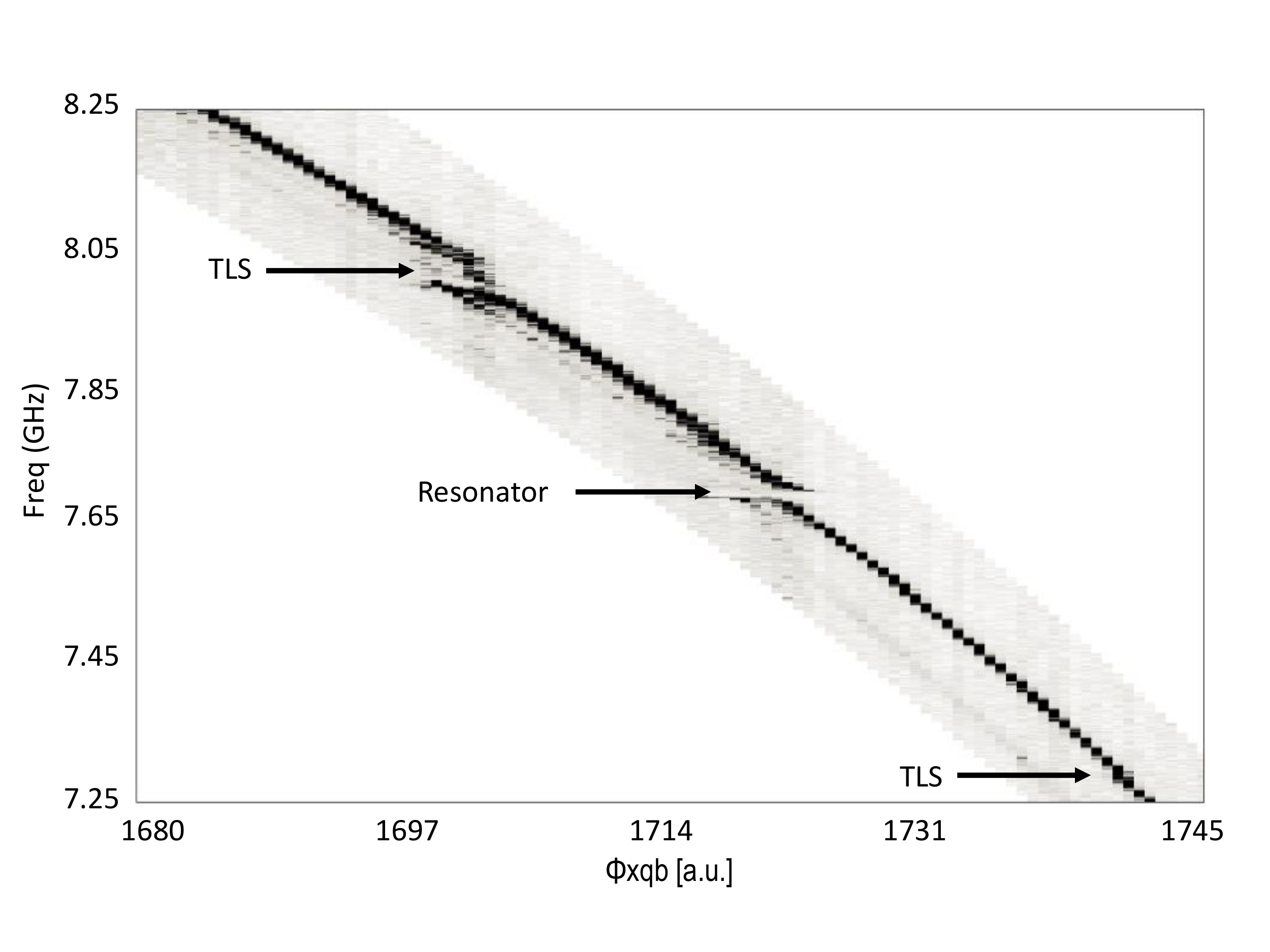}
\caption{Standard spectroscopy.  The arrows indicate avoided level
crossings or splittings due to coupling of the qubit to either an
engineered on-chip resonant cavity ($7.7 GHz$) or random spurious
TLS fluctuators.} \label{SpecFR3}
\end{figure}

A typical flux-biased phase qubit\cite{Simmonds} is composed of an
rf SQUID loop with critical current, $I_{q0}$, shunt capacitance,
$C_{qs}$, and geometric inductance, $L_q$.  The phase qubit,
described in more detail elsewhere\cite{Allman}, is coupled to
external control and readout circuitry. A dc bias line, coupled to
the qubit inductance via a mutual inductance, $M_{qb}$, provides
an external flux bias to the qubit. This bias controls the
non-linear Josephson inductance of the qubit which controls the
energy level spacing between qubit states as well as level
anharmonicity.  The qubit is operated in a flux bias regime that
creates an approximately cubic potential energy well of sufficient
anharmonicity to reliably isolate the two lowest metastable energy
states for qubit operations. A microwave drive, either
capacitively or inductively coupled to the qubit, provides the
excitation energy to drive transitions between the two lowest
qubit levels, labeled $\vert{g}\rangle$ and $\vert{e}\rangle$
respectively.  A fast ($\sim5\,ns$) measure pulse is then applied
to induce tunneling of the $\vert{e}\rangle$ state to the
adjacent, stable well\cite{Cooper}.  The state of the qubit is
then read out via a dc SQUID coupled to the qubit's geometric
inductance via a mutual inductance, $M_{qs}$.  The junctions are
$6.5\;\mu{m}^2$ via-style junctions with an rf plasma clean used
to remove the native oxide barrier before a room temperature
thermal oxidation.

In standard spectroscopic measurements\cite{Simmonds}, the excited
state probability is measured as a function of both drive
frequency and applied flux.  For a given bias flux, when the
applied microwave drive is on resonance with the qubit, the
excited state probability peaks.  When the qubit transition
frequency nears the resonant frequency of a TLS, an avoided
crossing occurs, splitting the resonant peak into two peaks
(Figure \ref{SpecFR3}).  The size of the splitting $S$ is a
measure of the coupling strength $hS/2$ between the qubit and the
TLS. The smaller the coupling strength, the smaller the splitting
size. Long excitation pulse times ($\sim 500\,ns$) and typical
qubit linewidths, on the order of $2-10\,MHz$, can limit the
ability to resolve the behavior of weakly coupled TLSs.

\begin{figure}[!htbp]
\centering
\includegraphics[width=\columnwidth]{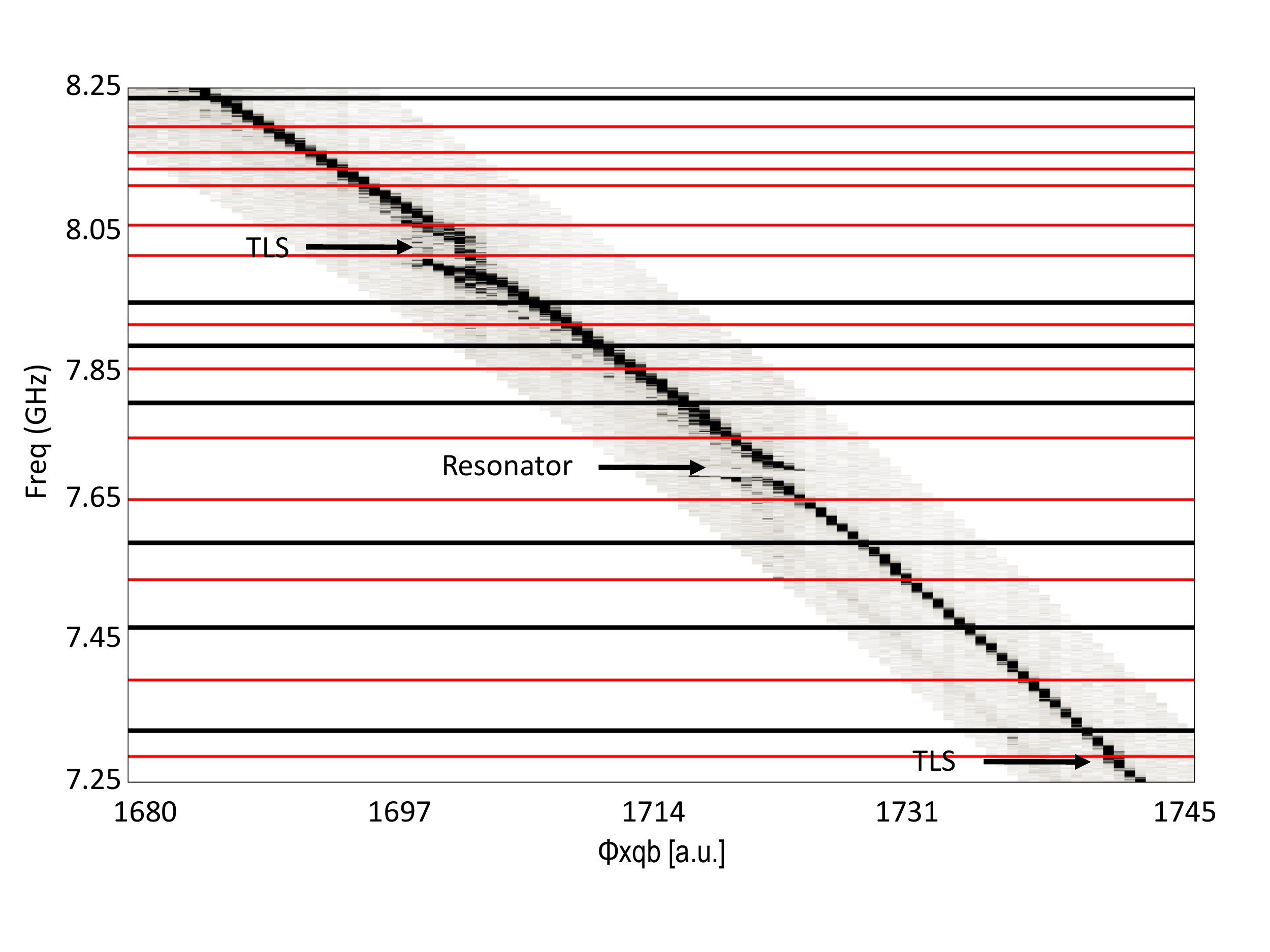}
\caption{The same spectroscopy from Figure \ref{SpecFR3} with
horizontal red and black lines indicating the qubit frequencies
where energy relaxation data (shown in Figure \ref{T1s}) was
obtained.} \label{T1Spec}
\end{figure}

Traditional spectroscopy scans are time-consuming. To achieve a
moderately high resolution scan, the step change on the frequency
axis is $\Delta{f}\sim{1\,MHz}$.  To capture the qubit's resonance
peak with reasonable detail, the total sweep width is typically
$\sim {200\,MHz}$. Along the applied flux axis, the resolution is
typically on the order of $1/2 m\Phi_0$ over a range of about $250
m\Phi_0$. The resulting total number of data points required for a
standard qubit spectroscopy is $N_t = N_fN_{\Phi}\approx{10^5}$.
The acquisition time per data point, governed by the number of
measurements per point as well as the ``dead'' time associated
with the instrument control software is $t_p\sim{0.5\,s}$.  Thus
the time to acquire a basic spectroscopy data set is $t_{spec} =
N_tt_p\approx{15\,hrs}$. Figure \ref{SpecFR3} shows a typical
phase qubit spectroscopy over a range of about $70 m\Phi_0$ in
applied qubit flux.  This particular device\cite{Allman} was
intentionally strongly coupled to a lumped element resonator as
indicated by the avoided crossing at $7.75\,GHz$.  The horizontal
arrows indicate avoided crossings due to coupling to TLS
fluctuators.

\begin{figure}[!htbp]
\centering
\includegraphics[width=\columnwidth]{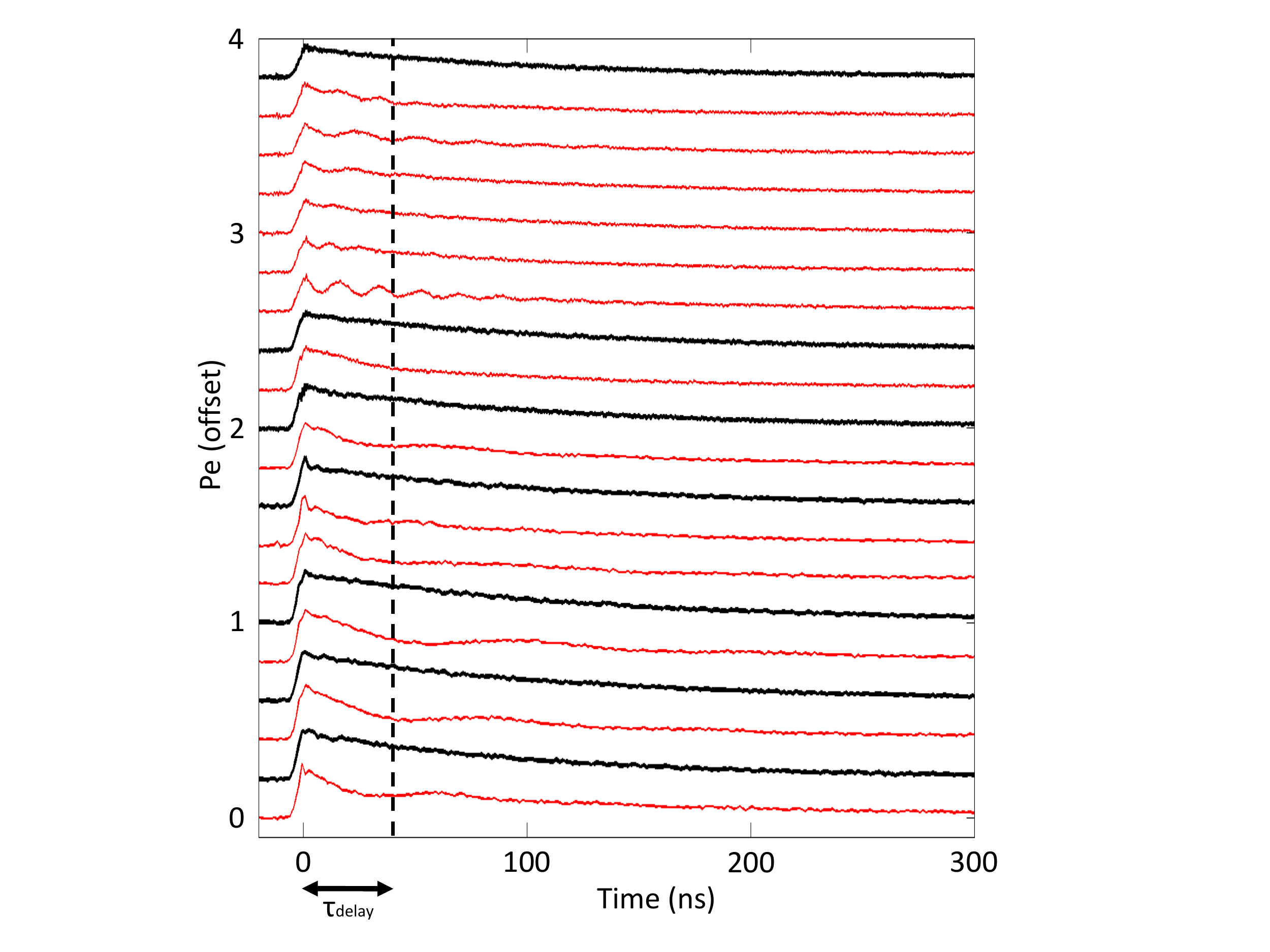}
\caption{Energy relaxation ($T_1$) measurements at particular bias
fluxes and frequencies identified in the spectroscopy of Figure
\ref{T1Spec}. The black curves show an exponential dependence as
expected. The red curves show coherent oscillations of varying
frequency.} \label{T1s}
\end{figure}

Another useful measurement is to determine the energy relaxation
time ($T_1$) of the qubit. This is done by applying a $\pi$-pulse
to the qubit and then sweeping the delay time $\tau_d$ between the
measure pulse and the end of the $\pi$-pulse. If the qubit is not
interacting with another system (other than the environmental bath
with many degrees of freedom), the excited state probability
decays exponentially in time. If this measurement is performed
while the qubit is on resonance with another quantum system, the
resultant curve will coherently oscillate with a period
proportional to the coupling strength between the qubit and the
other system.

In order to obtain a reliable measurement of the qubit's energy
relaxation time, the qubit should be far-detuned from any other
coupled systems including TLS fluctuators whose position has
previously been determined by visible splittings in the
spectroscopic data. According to Figure \ref{SpecFR3}, the energy
relaxation curves should be exponential as long as the qubit's
resonant frequency is within a clean region, at least a splitting
size away from the center of any splittings.

What we observe is illustrated in Figures \ref{T1Spec} and
\ref{T1s}. On resonance with any visible splittings, we find
coherent oscillations as observed previously with phase
qubits\cite{Cooper}. Remarkably, however, we also find many places
within the qubit's spectral range where the time domain data
yields coherent-like oscillations with no evidence of a splitting
in the corresponding spectroscopic measurements shown in Figure
\ref{T1Spec}. These oscillations also vary in frequency indicating
a random distribution of weak coupling strengths between the TLSs
and the qubit as found for larger coupling
strengths\cite{MartinisC}. The observation of these weakly coupled
TLS fluctuators is consistent with predictions based on the
standard TLS model for defects in amorphous dielectric
solids\cite{MartinisC}. The expected distribution of splitting
sizes given by Eq. 4 in Ref. \cite{MartinisC} shows that the
defect density scales approximately as $1/S$ where $S$ is the
splitting size (in $GHz$) and the coupling strength is given by
$hS/2$. Our measurements qualitatively agree with this prediction
$-$as the coupling strength decreases, the defect density
increases. The measurements recorded in Ref. \cite{MartinisC}
relied on traditional spectroscopic measurements with a minimum
splitting resolution of 10 $MHz$.

\begin{figure}[!htbp]
\centering
\includegraphics[width=\columnwidth]{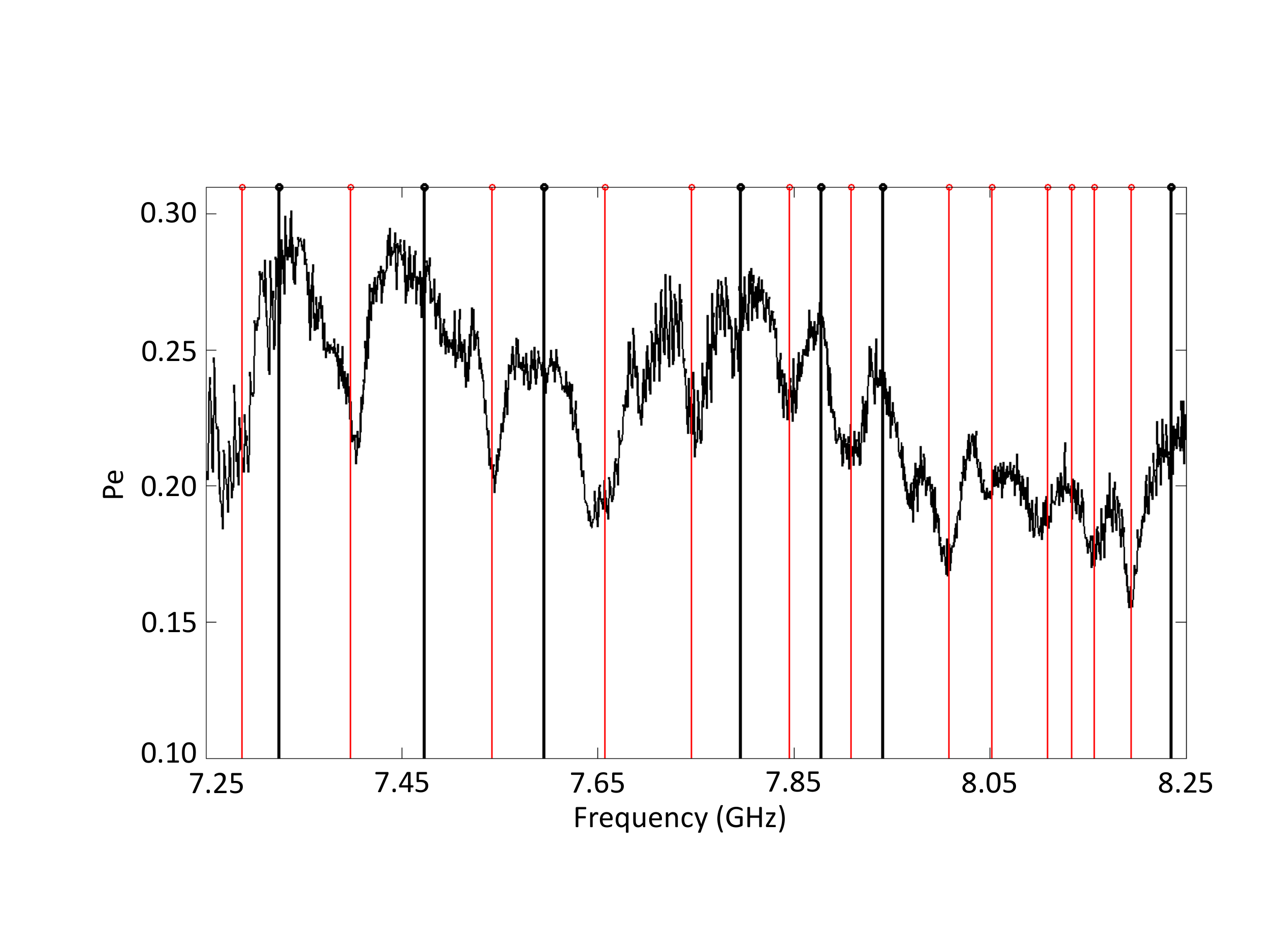}
\caption{A time-domain dip-scan showing higher spectral TLS
density than the standard spectroscopic scan.  The peaks
correspond to regions in the qubit spectroscopy where the $T_1$
decay curve is exponential.  The dips correspond to places where a
coherent oscillation is present, identifying a TLS fluctuator in
the qubit. Note that these dips occur where the standard
spectroscopy curve appears to be free from any TLS fluctuators.}
\label{AT1DS}
\end{figure}

We have devised a relatively rapid experimental technique for
locating the position of these weakly coupled ($S<10\,MHz$) TLS's
throughout the qubit's entire spectral range. Once standard
spectroscopy has been performed, we have a calibration of the
resonant frequency of the qubit as a function of bias flux. We can
now search for coherent oscillations at each qubit frequency.
Performing high resolution `$T_1$-scans' of time domain energy
relaxation measurements will certainly reveal the TLS features as
coherent oscillations but with data acquisition times that will be
as long as standard spectroscopy. In order to reduce the number of
data points for a given frequency range of the qubit, we choose a
different approach.  We hold the measure delay time $\tau_d$ fixed
at a particular value, just after the maximum excitation of the
qubit from the $\pi$-pulse. This value is a small fraction of the
energy relaxation time of the qubit, sampling a single point early
in the decay with nearly maximum probability. For a given flux, if
the qubit is free from interactions with any other systems, the
probability amplitude remains high. However, if the qubit is on
resonance with a TLS (or any other coherent system), the
probability amplitude with undergo oscillations that can produce a
`dip' in probability amplitude at the specific sampling point
chosen. By taking a single data point for each qubit frequency, we
have reduced the required number of points, spanning only the flux
dimension, allowing finer resolution `dip-scans' with fewer points
and hence shorter acquisition times.

Figure \ref{AT1DS} shows a dip-scan with $\tau_d=40\,ns$. Here the
resolution in applied qubit flux is $\sim 50 m\Phi_0$ for a total
of $1500$ data points with a corresponding acquisition time of
approximately $\sim20$ minutes. Notice that the dips in this scan
correspond directly with the TLS fluctuators identified in the
full time domain energy relaxation ($T_1$) curves shown in Figure
\ref{T1s}. It is evident that this technique allows us to count
the number of TLS fluctuators with higher resolution than the
standard spectroscopy shown in Figure \ref{SpecFR3}.

We have devised a new method for identifying TLS fluctuators in
superconducting phase qubits. This `dip-scan' method is general
purpose and can be applied to all superconducting qubits with a
tunable frequency. This method is useful for future
characterizations of Jospehson junction based qubits and may help
to elucidate the origin of TLS fluctuators, facilitate their
elimination, and eventually lead to increases in superconducting
qubit coherence times.

This work was supported by NIST and ARO Grant No.
W911NF-06-1-0384.

\bibliographystyle{apsrev4-1}

%\bibliography{Tunable}

%

\end{document}